\documentstyle[12pt]{article}

\def\be{\begin{equation}}
\def\ee{\end{equation}}
\def\bea{\begin{eqnarray}}
\def\eea{\end{eqnarray}}

\begin{document}
\title{S-Duality in Gauge Theories as a Canonical Transformation}
\author{\small
Y. Lozano,
\thanks{\tt yolanda@puhep1.princeton.edu
} \\
\small\it Joseph Henry Laboratories,\\
\small\it Princeton University,\\
\small\it Princeton, New Jersey 08544, USA}
\date{}
\maketitle
\setcounter{page}{0} \pagestyle{empty}
\thispagestyle{empty}
\vskip 1.0in
\begin{abstract}
We show that S-duality in four dimensional non-supersymmetric
abelian gauge theories can be
formulated
as a canonical transformation in the phase space of the theory.
This transformation is the usual interchange between electric and
magnetic degrees of freedom.
It is shown that in phase space the modular
anomaly emerges as the result of integrating out the momenta degrees
of freedom. The generalization to $d$ dimensional abelian gauge theories of
$p$-forms is also considered.
In the case of non-abelian gauge theories a careful analysis of the
constraints implied by the canonical transformation shows that it
does not relate Yang-Mills theories with inverted couplings. In fact
the dual theory is shown to be
of Freedman-Townsend's type, also with
${\tilde \tau}=-1/\tau$,
$\tau=\frac{\theta}{2\pi}+
\frac{4\pi i}{g^2}$.
\end{abstract}
\vfill
\begin{flushleft}
PUPT-1552\\
hep-th/9508021\\
August 1995
\end{flushleft}
\newpage\pagestyle{plain}

\def\theequation{\thesection . \arabic{equation}}

\section{Introduction}
\setcounter{equation}{0}

A lot of progress has been made in the last few years in the
understanding of S-duality as a symmetry of four dimensional gauge
theories.  The conjecture of Montonen and Olive \cite{mo} that $N=4$
supersymmetric Yang-Mills theories were invariant under strong-weak
coupling with the exchange of the gauge group by its dual was tested
in \cite{vw}, were it was shown that in fact the partition function
transformed as a modular form. Some progress has been also made for
$N=2$ and $N=1$ supersymmetric Yang-Mills theories \cite{sw,seiberg}.
However a path integral derivation of S-duality is in general still
unknown.  In \cite{witten} Witten showed that S-duality in four
dimensional abelian gauge theories \cite{cr,cardy,shw} can be
implemented at the level of the path integral in a very similar way to
T-duality in non-linear sigma models in String Theory \cite{reviews}.
The idea is to consider a global isometry of the Lagrangian which can
be expressed as translations of a given coordinate (the adapted
coordinate), gauge this isometry by introducing a fake gauge field and
impose the constraint that the curvature tensor associated to this
gauge field is zero so that the gauge field is non-propagating.
Integrating the Lagrange multiplier and fixing the gauge field to zero
the original theory is recovered and integrating the gauge field and
fixing the adapted coordinate to zero the new dual theory is obtained.
In the case of T-duality the initial variables are 0-forms and the
global isometry that is gauged is $\theta\rightarrow \theta+\epsilon$
where $\theta$ is the adapted coordinate. In the case of abelian gauge
theories the initial variables are 1-forms and the isometry which is
gauged is $A\rightarrow A+\epsilon$ where now the $\epsilon$ parameter
is a 1-form. Then the gauge field which has to be introduced is a
2-form and its field strength a 3-form. In 4 dimensions the Lagrange
multiplier imposing that the field strength vanishes is a 1-form, like
the original gauge field, and the dual theory is expressed also in
terms of 1-forms. Also for this non-supersymmetric case the partition
function transforms as a modular function with a modular weight
proportional to the Euler characteristic and the signature of the
manifold \cite{witten,verlinde}.

Given the analogy with T-duality a canonical transformation must be
beyond this path integral manipulation, since this is the case in
T-duality \cite{venezia,aagl2,la}.  In section 2 we present the
explicit generating functional producing this transformation and show
that it is the generalization of the functional in 2-dimensional
non-linear sigma models to 4 dimensions and 1-forms.  Under this
transformation electric and magnetic degrees of freedom get
interchanged (with the minus relative sign) as shown in abelian
lattice gauge theories in \cite{cardy}. The canonical transformation
approach is the simplest in order to obtain the dual theory, also in
this case in which in the Hamiltonian formulation one has to be
careful with the constraints. It is easy to show that both the initial
and the dual theory are defined in the same subspace of the phase
space after the canonical transformation is performed. We show that in
phase space the partition functions of the initial and dual theories
coincide and that only after integrating out the momenta degrees of
freedom the modular anomaly \cite{witten,verlinde} appears.

The same canonical transformation applied to the non-abelian case
seems to relate Yang-Mills theories with inverted couplings.
However a careful analysis of the constraints points out that this
is not the case.
The dual theory is in fact of Freedman-Townsend's
type \cite{ft}, i.e. it is expressed as a function of arbitrary 2-forms
which are not derived from a vector potential.
We show this in section 3.

The results presented in section 2 can be easily generalized to the
case of $d$ dimensional abelian gauge theories of $p$-forms, as it is
explained in section 4. The modular anomaly in the transformation of
the partition function is obtained. The implementation at the level of
the path integral using a coset construction was presented in
\cite{barbon}.

\section{The abelian case}
\setcounter{equation}{0}

In this section we construct the explicit canonical transformation
which produces the change
\be
\label{2uno}
\tau\rightarrow -1/\tau
\ee
with $\tau=\theta/2\pi+4\pi i/g^2$, for
$U(1)$ four dimensional euclidean gauge theories.

Let us consider the Lagrangian
\bea
\label{2dos}
L&=&\frac{1}{8\pi}(\frac{4\pi}{g^2}F_{mn}F^{mn}+\frac{i\theta}{4\pi}
\epsilon_{mnpq}F^{mn}F^{pq}) \nonumber\\
&=&\frac{i}{8\pi}({\bar \tau}
F^+_{mn}F^{+mn}
-\tau F^-_{mn}F^{-mn})
\eea
where
\bea
\label{2tres}
F^+_{mn}&=&\frac12 (F_{mn}+\,^*F_{mn})=\frac12
(F_{mn}+\frac12 \epsilon_{mnpq}F^{pq}),\nonumber\\
F^-_{mn}&=&\frac12 (F_{mn}-\,^*F_{mn})=\frac12
(F_{mn}-\frac12 \epsilon_{mnpq}F^{pq})
\eea
and $F_{mn}=\partial_m A_n-\partial_n A_m$.  It was shown in
\cite{witten} that the transformation (\ref{2uno}) could be derived at
the level of the path integral by the usual Rocek and Verlinde's
procedure \cite{rv} one follows to construct abelian T-duals of two
dimensional sigma models in String Theory. In this case given a global
abelian continuous isometry of the sigma model one can turn it local
by introducing a fake gauge field in the Lagrangian by minimal
coupling and imposing the constraint that this gauge field is
non-dynamical. Solving this constraint and fixing the gauge field to
be zero one recovers the original theory. If instead the gauge field
is integrated and the gauge is fixed in the original variables a sigma
model written in terms of the Lagrange multiplier introduced to impose
the constraint is obtained. This is the dual sigma model. In
\cite{witten} the same construction is applied to obtain the dual of
the four dimensional abelian gauge theory.
The global continuous abelian isometry in
this theory is
\be
\label{2cuatro}
A\rightarrow A+\epsilon
\ee
where now the isometry parameter is a 1-form. This global isometry can
be gauged by introducing a gauge field $G$, 2-form,
which is imposed to be non-dynamical with the term
\be
\label{2cinco}
\int_M d^4 x {\tilde A} dG
\ee
where the Lagrange multiplier ${\tilde A}$ is a 1-form. Integrating
${\tilde A}$ the
constraint $dG=0$ is obtained, ie. $G$ pure gauge, and we can recover
(\ref{2dos}) by either fixing $A=0$ or $G=0$. On the other hand by
integrating out $G$ and then fixing $A=0$ the following Lagrangian is
gotten:
\be
\label{2seis}
{\tilde L}=\frac{i}{8\pi}(-\frac{1}{{\bar \tau}} {\tilde F}^+_{mn} {\tilde
  F}^{+mn}
+\frac{1}{\tau} {\tilde F}^-_{mn}{\tilde F}^{-mn})
\ee
with ${\tilde F}^{\pm}$ the self- and antiself-dual components of
${\tilde F}_{mn}\equiv\partial_m {\tilde A}_n-\partial_n {\tilde A}_m$.
This is the S-dual of the initial electromagnetic theory since
in the particular case $\theta=0$ it corresponds to the inversion of
the coupling constant $g$.

In this procedure we have made an integration by parts in the Lagrange
multipliers term and neglected a total derivative\footnote{This term
  is seen in the gauge $A=0$.}. However this total derivative contains
some information, in particular it implies that the initial and dual
Lagrangians are equal up to a total time derivative, exactly what
happens when two theories are related by a canonical transformation.
To be more precise, the generating functional of a canonical
transformation from $\{q^i,p_i\}$ to $\{Q^i,P_i\}$ is such that
\be
\label{2siete}
p_i \dot{q^i}-H(q^i,p_i)=P_i \dot{Q^i}-\tilde{H}(Q^i,P_i)+
\frac{d{\cal F}}{dt}.
\ee
If ${\cal F}$ is a type I generating functional (depending only on coordinates)
$H=\tilde{H}$ if and only if\footnote{We assume ${\cal F}$ does not depend
explicitly on time.}
\bea
\label{2ocho}
&&\frac{\partial {\cal F}}{\partial q^i}=p_i \nonumber\\
&&\frac{\partial {\cal F}}{\partial Q^i}=-P_i
\eea
Under duality\footnote{We
have dropped
the global $i/8\pi$ factor. It will then appear when exponentiating
these quantities.}:
\be
\label{2nueve}
{\tilde L}({\tilde A})=L(A)+d{\tilde A}\wedge dA
\ee
which implies\footnote{Our convention for the product of forms is:
${\tilde F}\wedge F=\epsilon^{mnpq}{\tilde F}_{mn} F_{pq}$.}
\be
\label{2diez}
\epsilon^{mnpq} (\partial_m {\tilde A}_n-\partial_n {\tilde A}_m)
(\partial_p A_q-\partial_q A_p)=-(\frac{\delta {\cal F}}{\delta {\tilde A}_m}
\dot{{\tilde A}}_m
+\frac{\delta {\cal F}}{\delta A_m}\dot{A}_m)
\ee
This produces the canonical transformation
\bea
\label{2once}
&&{\Pi}^\alpha=\frac{\delta {\cal F}}{\delta A_\alpha}=
-4\,^*{\tilde F}^{0\alpha}
,\qquad \Pi^0=0, \nonumber\\
&&{\tilde \Pi}^\alpha=-\frac{\delta {\cal F}}{\delta {\tilde A}_\alpha}
=4\,^*F^{0\alpha},
\qquad {\tilde \Pi}^0=0
\eea
plus a constraint
\be
\label{2doce}
\Pi^\alpha\partial_\alpha A_0={\tilde \Pi}^\alpha\partial_\alpha
{\tilde A}_0,
\ee
where greek indices run over spatial coordinates.

The generating functional producing this canonical transformation
is
\be
\label{2trece}
{\cal F}=-2\int_{M, t fixed} d^3 x ({\tilde A}_\alpha\,^*F^{0\alpha}+
A_\alpha\,^*{\tilde F}^{0\alpha})=-\frac12 \int_M d^4 x
{\tilde F}\wedge F.
\ee
This is the result one would expect a priori from what is known in
two-dimensional sigma-models, where the generating functional is given in
terms of the adapted coordinate to the isometry $\theta$ and the Lagrange
multiplier ${\tilde \theta}$ by \cite{venezia,aagl2}
\be
\label{2catorce}
{\cal F}=-\frac12 \int_{M_2} d{\tilde \theta}\wedge d\theta.
\ee
The Hamiltonian associated to (\ref{2dos}) is given by:
\be
\label{2quince}
H=\frac{1}{4(\bar{\tau}-\tau)}\Pi_\alpha \Pi^\alpha+\partial_\alpha A_0
\Pi^\alpha-\frac{\bar{\tau}+\tau}{\bar{\tau}-\tau}\Pi_\alpha
\,^*F^{0\alpha}
+\frac{4\bar{\tau}\tau}{\bar{\tau}-\tau}\,^*F^{0\alpha}
\,^*F_{0\alpha}
\ee
plus the constraints
\be
\label{2dieciseis}
\Pi_0=0,\qquad \partial_\alpha \Pi^\alpha=0,
\ee
where
\be
\label{2dieciseisbis}
\Pi^\alpha=
4{\bar \tau} F^{+0\alpha}-4\tau F^{-0\alpha}.
\ee
$\Pi_0$ is a primary constraint and $\partial_\alpha \Pi^\alpha=0$
is the secondary constraint emerging from the equation of motion for
$\Pi_0$. They imply that the theory is defined in the reduced phase
space
given by $\Pi_0=0$, $\partial_\alpha \Pi^\alpha=0$.
These constraints are also satisfied in the dual theory,
since they are obtained directly
from the canonical transformation. Then the dual theory is
defined in the same reduced phase space than the original one.
The relation (\ref{2doce}) is
trivial in this subspace. However we need to consider it in
order
to recover the dual Lagrangian from the canonically transformed
Hamiltonian, since for
that we need the naive Hamiltonian without taking into
account the constraints. Our purpose is to show that the
canonically
transformed Lagrangian is the dual Lagrangian and for that we do not
need to study
in detail the way the theory gets defined in the Hamiltonian formalism
\cite{Ramond}, it is enough to show that both the initial and dual
theories
are defined in the same reduced phase space.

The canonically transformed Hamiltonian reads:
\be
\label{2diecisiete}
{\tilde H}=\frac14 \frac{\bar{\tau}\tau}{\bar{\tau}-\tau}
{\tilde \Pi}_\alpha
{\tilde \Pi}^\alpha+\partial_\alpha {\tilde A}_0 {\tilde \Pi}^\alpha+
\frac{\bar\tau
+\tau}{\bar{\tau}-\tau}{\tilde \Pi}_\alpha\,^*{\tilde F}^{0\alpha}+
\frac{4}{\bar{\tau}-\tau}\,^*{\tilde F}_{0\alpha}
\,^*{\tilde F}^{0\alpha}.
\ee
The corresponding Lagrangian is given by the dual
Lagrangian (\ref{2seis}). Recall that (\ref{2once}):
\bea
\label{217a}
&&\Pi^\alpha=-4\,^*{\tilde F}^{0\alpha}, \nonumber\\
&&{\tilde \Pi}^\alpha=4\,^*F^{0\alpha} \nonumber
\eea
corresponds to the usual interchange between electric and magnetic
degrees of freedom when there is no $\theta$-term.

Some useful information can be obtained within this approach.
The generating functional (\ref{2trece}) is linear in both the original
and dual variables. Then the following relation holds:
\be
\label{217b}
H e^{\frac{i{\cal F}}{8\pi}}={\tilde H}e^{\frac{i{\cal F}}{8\pi}}
\ee
which implies:
\be
\label{217c}
\psi_k[{\tilde A}]=N(k)\int {\cal D}A(x^\alpha)
e^{\frac{i}{8\pi}{\cal F}[{\tilde
    A},A(x^\alpha)]}
\phi_k[A(x^\alpha)]
\ee
with $\phi_k[A]$ and $\psi_k[{\tilde A}]$ eigenfunctions of the
initial
and dual Hamiltonians respectively with the same eigenvalue
and $N(k)$ a normalization factor \cite{ghandour}.
{}From this relation global properties can be easily worked out.
The Dirac quantization condition:
\be
\label{217d}
\int_{\Sigma}F=2\pi n,\quad n\in Z,
\ee
for $\Sigma$ any closed two-surface in the manifold,
implies for ${\tilde F}$:
\be
\label{217e}
\int_{\Sigma}{\tilde F}=2\pi m,\quad m\in Z
\ee
and ${\tilde F}$ must live in the dual lattice.
Also from (\ref{217c}) the
transformation applies to any four dimensional manifold $M$ since
$\phi_k[A]$ can be the result of integrating the theory in an
arbitrary manifold with boundary.

We can obtain in phase space the modular anomaly emerging in the
transformation of the partition function \cite{witten,verlinde}.
The argument goes
as follows. In phase space the partition function is given
by\footnote{In order to have a well-defined partition function we have
to fix the gauge symmetry. The following arguments are in this
sense formal.}:
\be
\label{2dieciocho}
Z_{ps}=\int {\cal D}A_\alpha {\cal D}\Pi^\alpha e^{-
\frac{i}{8\pi}\int d^4x (\dot{A}_\alpha
\Pi^\alpha-H)}
\ee
Under (\ref{2once})
\be
\label{218b}
{\cal D}A_\alpha {\cal D}\Pi^\alpha={\cal D}{\tilde A}_\alpha {\cal D}
{\tilde \Pi}^\alpha.
\ee
Then the dual phase space partition
function is given by:
\be
\label{2diecinueve}
{\tilde Z}_{ps}=\int {\cal D}{\tilde A}_\alpha
{\cal D}{\tilde \Pi}^\alpha
e^{-\frac{i}{8\pi}\int d^4x (\dot{{\tilde A}}_\alpha
{\tilde \Pi}^\alpha-{\tilde H})}=Z_{ps}
\ee
showing that in phase space the partition function is invariant
under duality.
Integration on momenta in (\ref{2dieciocho}) gives:
\be
\label{2veinte}
Z_{ps}=\int {\cal D}A_\alpha ({\rm Im} \tau)^{B_2/2} e^{-\int d^4x L}
\ee
with $L$ given by (\ref{2dos}). The factor $({\rm Im} \tau)^{B_2/2}$ in the
measure is the regularized $({\rm det\,Im}\tau)^{1/2}$ coming from the
gaussian integration over the momenta. $B_2$ is the dimension of the
space of
2-forms in the four dimensional manifold $M$ (regularized on
a lattice) and emerges because the momenta are 2-forms.

The same calculation in the dual phase space partition function gives:
\be
\label{2veintiuno}
{\tilde Z}_{ps}=\int {\cal D}{\tilde A}_\alpha ({\rm det}
({\rm Im} -\frac{1}{\tau}))^{1/2}
e^{-\int d^4x {\tilde L}}
\ee
with ${\tilde L}$ given by (\ref{2seis}). We regularize the factor
\be
\label{2veintidos}
({\rm det}({\rm Im} -\frac{1}{\tau}))^{1/2}=({\rm det}({\rm Im}
\tau /(\tau {\bar\tau})))^{1/2}
\ee
by
\be
\label{2veintitres}
({\rm Im} \tau)^{B_2/2}\bar\tau^{-B_2^+/2}\tau^{-B_2^-/2}
\ee
where $B_2^+$ and $B_2^-$ are respectively the dimensions of the
spaces
of self-dual and
anti-self-dual 2-forms.
In configuration space the partition function is defined by \cite{witten}:
\be
\label{2veinticuatro}
Z=({\rm Im} \tau)^{(B_1-B_0)/2}\int {\cal D}A_\alpha e^{-S}=
({\rm Im} \tau)^{(B_1-B_0-B_2)/2} Z_{ps}
\ee
and in the dual model
\be
\label{2veinticinco}
{\tilde Z}=(\frac{{\rm Im} \tau}{\tau \bar\tau})^{(B_1-B_0)/2}
\int {\cal D}
{\tilde A}_\alpha e^{-{\tilde S}}.
\ee
{}From $Z_{ps}={\tilde Z}_{ps}$ we arrive to
\be
\label{2veintiseis}
Z=\tau^{-(\chi-\sigma)/4}{\bar\tau}^{-(\chi+\sigma)/4} {\tilde Z}
\ee
where $\chi=2(B_0-B_1)+B_2$ is the Euler number (the regularization is
such
that $B_p=B_{4-p}$) and $\sigma=B_2^+-B_2^-$ is
the signature of the manifold. This is the modular factor appearing in
\cite{witten,verlinde}. In phase space the partition function is simply
defined as the integration over coordinates and momenta and it transforms
as a scalar with modular weight equal to zero. Is only when going to the
configuration space that the integrations over the momenta produce some
determinants which after being regularized yield the modular factor
found in \cite{witten,verlinde}. A very similar argument should apply
to the transformation of the dilaton in two-dimensional non-linear
sigma-models.

\section{The non-abelian case}
\setcounter{equation}{0}

The canonical transformation approach can be
generalized to the case
of
non-abelian gauge theories with compact group $G$.
The initial Lagrangian is given by:
\bea
\label{4uno}
L&=&\frac{1}{8\pi}(\frac{4\pi}{g^2}F^{(a)}_{mn} F^{(a)mn}+
\frac{i\theta}{4\pi}\epsilon^{mnpq}F^{(a)}_{mn} F^{(a)}_{pq})
\nonumber\\
&=&\frac{i}{8\pi}({\bar \tau} F^{(a)+}_{mn} F^{(a)+ mn}-\tau
F^{(a)-}_{mn}F^{(a)- mn})
\eea
where $F=dA-A\wedge A$ and we have chosen $Tr(T^a T^b)=\delta^{ab}$
($T^a$ are the generators of the Lie algebra). The conjugate momenta
and the Hamiltonian are:
\bea
\label{i1}
&&\Pi^{a\alpha}=
2({\bar \tau}-\tau)F^{(a)0\alpha}+2({\bar \tau}+\tau)\,
^*F^{(a)0\alpha} \nonumber\\
&&\Pi^{a0}=0
\eea
\be
\label{i2}
H=\frac14 \frac{1}{{\bar \tau}-\tau}\Pi^a_\alpha \Pi^{a\alpha}+
(\partial_\alpha A^a_0+f_{abc}A^b_0A^c_\alpha)\Pi^{a\alpha}-
\frac{{\bar \tau}+\tau}{{\bar \tau}-\tau}\Pi^{a\alpha}\,
^*F^{(a)}_{0\alpha}+\frac{4{\bar \tau}\tau}{{\bar \tau}-\tau}
\,^*F^{(a)}_{0\alpha}\,^*F^{(a)0\alpha},
\ee
with $f_{abc}$ the structure constants of the Lie algebra.
The equations of motion of the primary constraints $\Pi^{a0}=0$
imply:
\be
\label{41bf}
\partial_\alpha \Pi^{a\alpha}-f_{abc}A^b_\alpha\Pi^{c\alpha}=0,
\ee
so that we can ignore the second term in the Hamiltonian
keeping in mind that the theory is defined in the reduced phase space
given by the constraints.

In the non-abelian case it proves more useful to use
$\{\,^*F_{0\alpha},\Pi^\alpha\}$ as the coordinates in phase space and
look for a canonical transformation
\be
\label{41bb}
\{\,^*F_{0\alpha},\Pi^\alpha\}\rightarrow \{\,^*{\tilde F}_{0\alpha},
{\tilde \Pi}^\alpha\}.
\ee
Then in order to define correctly the phase space of the theory we
have to introduce first order formalism for the initial Lagrangian.
The idea is to introduce a Lagrangian $L[{\tilde F},A]$, where now
${\tilde F}$ are arbitrary two-forms in the manifold, arranged to
give ${\tilde F}=dA-A\wedge A$ from its equations of motion.
Now the ${\tilde F}$ have no dynamical meaning since they have no
time derivative, and the momenta are conjugate to the $A$-variables:
\be
\label{cor1}
\Pi^{am}=\frac{\delta L[{\tilde F},A]}{\delta {\dot A}^a_m}.
\ee

It is easy to see that the following Lagrangian:
\be
\label{cor2}
L[{\tilde F},A]=\frac{i}{8\pi}Tr(-\frac{1}{{\bar \tau}}
{\tilde F}^+_{mn}{\tilde F}^{+mn}+\frac{1}{\tau}
{\tilde F}^-_{mn}{\tilde F}^{-mn}-2({\tilde F}^+_{mn}F^{+mn}-
{\tilde F}^-_{mn}F^{-mn})),
\ee
with $F=dA-A\wedge A$, is such that (\ref{4uno}) is obtained when
solving the equations of motion for ${\tilde F}$.
The canonical momenta are given by:
\bea
\label{cor3}
&&\Pi^{a\alpha}=\frac{\delta L[{\tilde F},A]}{\delta {\dot A}^a_\alpha}
=-4\,^*{\tilde F}^{(a)0\alpha} \nonumber\\
&&\Pi^{a0}=0
\eea
and coincide with (\ref{i1}) when substituting the equations of
motion. The Hamiltonian is also given by (\ref{i2}).

The canonical transformation
\bea
\label{41bc}
&&\Pi^{a\alpha}=-4\,^*{\tilde F}^{(a)0\alpha} \nonumber\\
&&{\tilde \Pi}^{a\alpha}=4\,^*F^{(a)0\alpha},
\eea
i.e. the usual interchange between electric and magnetic degrees of
freedom, produces the following ``dual'' Hamiltonian:
\be
\label{corr1}
{\tilde H}=\frac14 \frac{{\bar \tau}\tau}{{\bar \tau}-\tau}
{\tilde \Pi}^{a\alpha}{\tilde \Pi}^a_\alpha+
\frac{{\bar \tau}+\tau}{{\bar \tau}-\tau}{\tilde \Pi}^{a\alpha}
\,^*{\tilde F}^{(a)}_{0\alpha}+\frac{4}{{\bar \tau}-\tau}
\,^*{\tilde F}^{(a)0\alpha}\,^*{\tilde F}^{(a)}_{0\alpha},
\ee
which is of the same form than the Hamiltonian of the initial
theory with ${\tilde \tau}=-1/\tau$. However one must be careful
with the constraints. In particular the secondary constraints
(\ref{41bf}) imply for the dual theory:
\be
\label{cor8}
\partial_\alpha\,^*{\tilde F}^{(a)0\alpha}-
f_{abc}A^b_\alpha ({\tilde F})
\,^*{\tilde F}^{(c)0\alpha}=0.
\ee
These equations are not satisfied by the Yang-Mills theory defined
from ${\tilde F}$, so although (\ref{corr1}) would naively imply
that the dual theory is a Yang-Mills theory with
${\tilde \tau}=-1/\tau$, the analysis of the constraints shows that
this is not the case. For the abelian theory the corresponding
equation implies that ${\tilde F}$ is defined from a dual vector
potential ${\tilde A}$, but the absence of a non-abelian analogue
of Poincar\`e's lemma does not allow to conclude the same in the
non-abelian case. Inversely if we would consider (\ref{cor8})
as an equation determining $A({\tilde F})$ we would find
incompatibility with (\ref{41bc}).
We can then conclude that the usual interchange between electric
and magnetic degrees of freedom does not relate Yang-Mills
theories with inverted couplings.

Let us now obtain the ``true'' dual theory. The point is to
realize that in (\ref{cor2}) we can integrate $A$ instead of
${\tilde F}$ and in this way a new theory is obtained.
The equations of motion for $A$ are:
\be
\label{cor5}
\partial_n \,^*{\tilde F}^{(a)mn}-f_{abc} A^b_n\,^*{\tilde F}^{(c)mn}
=0,
\ee
which imply:
\be
\label{ultima}
A^a_m=R^{ab}_{mn}\partial_p
\,^*{\tilde F}^{(b)np},
\ee
where $R$ is the inverse of ${\rm ad}\,^*{\tilde F}$ and it is a well
defined matrix for arbitrary ${\tilde F}$ in four dimensions.
Substituting in (\ref{cor2}) we get:
\be
\label{cor9}
{\tilde L}=\frac{i}{8\pi} (-\frac{1}{{\bar \tau}}
{\tilde F}^{(a)+}_{mn}{\tilde F}^{(a)+mn}+\frac{1}{\tau}
{\tilde F}^{(a)-}_{mn}{\tilde F}^{(a)-mn}+
2 R^{ab}_{mn}(\,^*{\tilde F})\partial_q\,^*{\tilde F}^{(a)qm}
\partial_p\,^*{\tilde F}^{(b)np}),
\ee
where ${\tilde F}$ are arbitrary two-forms in the manifold\footnote{
We have not written explicitly the determinant coming from the
gaussian integration.}. This
Lagrangian is of Freedman-Townsend's type \cite{ft}, i.e.
${\tilde F}$ is the ``fundamental'' variable, not defined
from a vector potential ${\tilde A}$, and it has been already
shown to be dual to Yang Mills \cite{hfs,moha,gs}. We should
remark that in order to have a well defined dual theory a
prescription must be given for the pole when $R$ becomes singular
\cite{gs}.
The Hamiltonian associated to (\ref{cor9}) looks quite different
from the Hamiltonian of Yang-Mills and it is not easy to see if
they could be related by a canonical transformation.
In any case such a transformation would have to map a
phase space of two-forms $\{\,^*F_{0\alpha},\Pi^\alpha\}$ into a
phase space of two and one forms $\{{\tilde F},{\tilde \Pi}\}$
(since the canonical momenta are conjugate to two forms),
which means that if existing at all it would be quite
different from an
interchange between electric and magnetic degrees of freedom.
Let us point out however that
the first equation in (\ref{41bc}) (which maps two-forms on
two-forms) produces the mapping from the
constraint (\ref{41bf}) of the initial theory to the equation
of motion (\ref{cor8})
of the dual. This means that after all this interchange could have
some physical meaning.

\vspace*{0.5cm}

In this section we have been able to obtain the dual theory
by manipulating the path integral in a similar way to the one
known in the abelian case.
One just needs to consider the intermediate Lagrangian
$L[{\tilde F},A]$ and integrate out either ${\tilde F}$ or $A$ to
obtain the initial or the dual Lagrangians.

The group in which the dual variables live is the dual of
the original group
(the metric defined by its weight vectors is the inverse of the one in the
original
gauge group). In order to see this one needs to proceed more carefully
in
the previous derivation and take the original gauge fields in the fundamental
representation of the gauge Lie algebra and the metric defined by the
weight vectors $g_{ab}\equiv d_{ab}$. Following the steps explained
above one arrives to a dual Lagrangian with metric
${\tilde g}_{ab}={\tilde d}_{ab}$ where ${\tilde d}_{ab}d^{bc}\equiv
\delta^c_a$.

\section{Generalization to p-forms abelian gauge theories}
\setcounter{equation}{0}

The generalization to $p$-forms abelian gauge theories in
$d$ dimensions is direct
from what we have studied in section 2\footnote{Dualization of
spin-0 and spin-2 gravity-like theories has been studied in \cite{curt}.}.
We are going to
consider the case $d=2(p+1)$ which is the one in which both the
initial
and dual theories are expressed as functions of
$(p+1)$-forms\footnote{In the
arbitrary case the dual theory would depend on $(d-p-1)$ forms.}.
The generalized S-duality transformation is implemented in the path
integral
by gauging the global isometry
\be
\label{3uno}
A\rightarrow A+\epsilon
\ee
where now $A$ and the gauge parameter are $p$-forms. The total derivative
term that gives information about the generating functional of the
canonical transformation is $d{\tilde A}\wedge dA$, with
${\tilde A}$, the Lagrange
multiplier, also a $p$-form.

It is immediate to show that the canonical transformation is
generated by the type-I generating functional\footnote{Our conventions
  are: $\,^*F^{i_1\dots i_{p+1}}=\frac{1}{(p+1)!}\epsilon^{i_1\dots
    i_d}
F_{i_{p+2}\dots i_d}$ and ${\tilde F}\wedge F=\epsilon^{i_1\dots i_d}
{\tilde F}_{i_1\dots i_{p+1}}F_{i_{p+2}\dots i_d}$.}:
\be
\label{3dos}
F=-\frac{1}{(p+1)!}\int d^dx d{\tilde A}\wedge dA
\ee
which produces:
\be
\label{3tres}
\Pi^{\alpha_1\ldots\alpha_p}=\frac{\delta F}{\delta
A_{\alpha_1\ldots\alpha_p}}=
-((p+1)!)^2\,^*{\tilde F}^{0\alpha_1\ldots\alpha_p}
\ee
\be
\label{3cuatro}
{\tilde \Pi}^{\alpha_1\ldots\alpha_p}=-\frac{\delta F}{\delta
{\tilde A}_{\alpha_1\ldots\alpha_p}}=
((p+1)!)^2\,^*F^{0\alpha_1\ldots\alpha_p}
\ee
The same relation (\ref{217c}) for the wave functionals holds in this
case since the generating functional is linear in the initial and dual
variables. From it we can obtain global information about the dual
variables.
We can also obtain the modular weight appearing in the transformation
of the partition function \cite{barbon}.

Let us consider first the case $p$ odd. $p+1$ is even and then the
theory
allows for a $\theta$-term. In phase space (we omit the $p$ indices):
\be
\label{3cinco}
Z_{ps}=\int {\cal D}A {\cal D}\Pi e^{-\frac{i}{8\pi}
\int d^dx({\dot A}\Pi-H)}=
({\rm Im}\tau)^{B_{p+1}/2}\int {\cal D}A e^{-S},
\ee
after regularizing the determinant coming from the gaussian
integration on
the momenta, $(p+1)$-forms in this case. The dual phase space partition
function coincides with the initial one and it is given by:
\be
\label{3seis}
{\tilde Z}_{ps}=\int {\cal D}{\tilde \Pi}{\cal D}{\tilde A}
e^{-\frac{i}{8\pi}\int d^dx ({\dot {\tilde A}}{\tilde \Pi}-{\tilde H})}
=({\rm Im}\tau)^{B_{p+1}/2} \tau^{-B^-_{p+1}/2}
{\bar \tau}^{-B^+_{p+1}/2} \int {\cal D}{\tilde A} e^{-{\tilde S}}
\ee
The configuration space partition function is:
\be
\label{3siete}
Z=({\rm Im}\tau)^{N_p/2}\int {\cal D}A e^{-S}
\ee
where we have followed the notation in \cite{barbon}, $N_p$ being the
dimension
of the space of $p$ forms after substracting all the gauge invariances
(see \cite{barbon} for the detailed analysis). In the dual model the
partition function is the
same with $\tau\rightarrow -1/\tau$. Then we have:
\be
\label{3ocho}
Z=\tau^{-\frac{\chi-\sigma}{4}}{\bar \tau}^{-\frac{\chi+\sigma}{4}}
{\tilde Z}
\ee
where $\chi=2(-1)^p N_p+(-1)^{p+1}B_{p+1}$ is the Euler number and
$\sigma=B^+_{p+1}-B^-_{p+1}$ the signature of the manifold.

In the case $p$ even a $\theta$-term does not exist. Similar
arguments to the ones above yield:
\be
\label{3nueve}
Z=(\frac{4\pi}{g^2})^{\chi/2}{\tilde Z}
\ee

All these results agree with the ones presented in \cite{barbon}.

\section{Conclusions}
\setcounter{equation}{0}

We have seen that for non-supersymmetric abelian four dimensional
gauge theories
S-duality can be
implemented
as a canonical transformation in the phase space of the theory which
is the usual interchange between electric and magnetic degrees
of freedom. This
transformation can be generalized to the case of non-abelian
gauge theories, where it seems to yield a Yang-Mills theory
with inverted couplings. However in this case the canonical transformation
produces some constraints in the dual theory which in the absence
of a non-abelian analogue of Poincar\`e's lemma do not imply that
a dual vector potential exists. The dual theory is not a
Yang-Mills theory but rather a
Freedman-Townsend's
type of theory \cite{ft} with inverted couplings.
This is shown by defining an intermediate Lagrangian depending on
the initial vector potential $A$ and the dual variables ${\tilde F}$
and from which the initial and dual Lagrangians are obtained
by integrating ${\tilde F}$ or $A$ respectively. It is argued that
if a canonical transformation relating the two theories exists it
would be far different from an interchange between electric and
magnetic degrees of freedom.

For the abelian case we have seen that in phase space the
partition function is invariant
under
S-duality and it is only after integrating out the momenta degrees of
freedom
that a modular factor appears and the partition function in
configuration
space transforms as a modular function.

We have generalized the canonical transformation approach to
$d$-dimensional
abelian
gauge theories defined with $p$ forms and obtained the
corresponding modular weights appearing in the transformation of the
partition function.

It could be very interesting to generalize the results presented in this
paper to the case of supersymmetric gauge theories.

\subsection*{Acknowledgements}

I would like to thank O. Alvarez and N. Mohammedi
for useful discussions and especially J.L.F. Barb\'on for
interesting
remarks leading to the final form of this paper.
A Fellowship from M.E.C. (Spain)
is acknowledged for partial financial support.

\end{document}